\documentclass[amsmath, amssymb, aps, pr d, reprint, longbibliography]{revtex4-2}
\usepackage{natbib}

\usepackage{hyperref}
\usepackage{graphicx}
\usepackage{url}
\usepackage{xcolor}
\usepackage{graphicx}
\usepackage{tabularx}
\usepackage{scalerel}
\usepackage{tikz}
\usetikzlibrary{svg.path}

\definecolor{orcidlogocol}{HTML}{A6CE39}
\tikzset{
  orcidlogo/.pic={
    \fill[orcidlogocol] svg{M256,128c0,70.7-57.3,128-128,128C57.3,256,0,198.7,0,128C0,57.3,57.3,0,128,0C198.7,0,256,57.3,256,128z};
    \fill[white] svg{M86.3,186.2H70.9V79.1h15.4v48.4V186.2z}
                 svg{M108.9,79.1h41.6c39.6,0,57,28.3,57,53.6c0,27.5-21.5,53.6-56.8,53.6h-41.8V79.1z M124.3,172.4h24.5c34.9,0,42.9-26.5,42.9-39.7c0-21.5-13.7-39.7-43.7-39.7h-23.7V172.4z}
                 svg{M88.7,56.8c0,5.5-4.5,10.1-10.1,10.1c-5.6,0-10.1-4.6-10.1-10.1c0-5.6,4.5-10.1,10.1-10.1C84.2,46.7,88.7,51.3,88.7,56.8z};
  }
}

\newcommand\orcidicon[1]{\href{https://orcid.org/#1}{\mbox{\scalerel*{
\begin{tikzpicture}[yscale=-1,transform shape]
\pic{orcidlogo};
\end{tikzpicture}
}{|}}}}

\usepackage{hyperref}

\usepackage{bm}
\usepackage{xcolor}
\usepackage{bm}
\usepackage{slashed}
\usepackage{xparse}
\usepackage{dsfont}
\usepackage{float}
\usepackage{etoolbox}
\patchcmd{\section}
  {\centering}
  {\raggedright}
  {}
  {}
\patchcmd{\subsection}
  {\centering}
  {\raggedright}
  {}
  {}
\DeclareUnicodeCharacter{2212}{-}
\allowdisplaybreaks
\begin{document}
\title{Ultraviolet dimensional reduction of spacetime with zero-point length}
\author{Vikramaditya Mondal \orcidicon{0000-0002-8565-5334}}%
 \email{vikramaditya.academics@gmail.com}
 \address{School of Physical Sciences, Indian Association for the Cultivation of Science, Kolkata, West Bengal 700032}

\begin{abstract}
    Among the many disparate approaches towards quantum gravity, the reduction of spacetime dimension in the ultraviolet regime is expected to be a common thread. The spectral dimension of spacetime is defined in the context of diffusion processes on a manifold. We show that a spacetime with zero-point length has spectral dimension $3.5$ when the characteristic diffusion time has the size of the zero-point length. The spectral dimension (negatively) diverges for an even shorter diffusion time, thus preventing reliable physical interpretation in the deep ultraviolet regime. The thermodynamic dimension is defined by realizing that the free energy $F(\beta)$ of a free field or ideal gas at finite temperature ($\beta^{-1}$) in $D$ dimensions scales as $F\sim \beta^{-D}$. Using Schwinger's proper time formalism, we show that for spacetime incorporating a zero-point length, the thermodynamic dimension reduces to $1.5$ near the Planck scale and then to $1$ in the deep ultraviolet regime. This signifies a “phase-transition” in which a (massless) bosonic ideal gas in four dimensions essentially behaves like radiation ($w=1/3$) at low energies, whereas near the Planck scale, it behaves equivalently to having an equation of state parameter $w=2$. Furthermore, dimension can be deduced from the potential $V_D(r)$ of interaction between two point-like sources separated at a distance $r$ as its scaling depends on $D$. Comparing with the scaling behavior of conventional Yukawa-like potentials at short distances, we show that the ultraviolet dimension appears to be either $2$ or $3$ depending on the use of massive or massless force carriers as probes.
\end{abstract}

\maketitle
\section{Introduction}
Without a complete or unique quantum theory of gravitation, it has become essential to understand what we can expect in the ultraviolet (corresponds to high energies) regime regarding the nature of spacetime, which is supposed to acquire quantum properties. In ``traditional'' classical physics, we have an operational view of spacetime as an arrangement of ideal clocks and rulers. The dimension of such spacetime is a discrete integer. However, there is growing evidence that in the ultraviolet regime, such a picture has to be abandoned as the dimensionality of the spacetime in such high energies may effectively deviate from its infrared (corresponds to low energies) value \cite{carlip2017dimension}.\par
Fractals are mathematical objects with a different dimension than the space they are part of due to their different scaling property at smaller scales, often resulting from self-similarity. For this reason, the deviation of the spacetime dimension from its conventional value is often interpreted as spacetime having a fractal-like structure at small scales. In many approaches to quantum gravity, the spacetime is viewed to have a \textit{grainy} or \textit{fuzzy} structure at small scales. In such a picture, if the graininess has a progressively self-similar structure at smaller scales, then the fractal interpretation becomes natural (see, for example, \cite{modesto2010spectral}). Among many conceptually disparate approaches towards quantum gravity, the replacement of the concept of dimension as a discrete integer with a dynamical variable that continuously changes or `flows' from its conventional counterpart at infrared regime to a different value in the ultraviolet regime seems to have a ubiquitous presence, for example, it is present in---String theory \cite{atick1988hagedorn,PhysRevD.56.2352,aurilia2002fuzzy}, Causal Dynamical Triangulation \cite{ambjorn2005spectral}, Causal Sets \cite{carlip2015dimensional}, Loop Quantum Gravity \cite{modesto2008fractal,modesto2009fractal}, Noncommutative spacetime \cite{benedetti2009fractal,nozari2015high}, Ho\v rava-Lifshitz gravity \cite{hovrava2009spectral}, Asymptotic Safety \cite{niedermaier2006asymptotic}, Super-renormalizable models of quantum gravity \cite{modesto2012super,modesto2017nonlocal,briscese2020unattainability}, models with `evolving' or `vanishing' dimensions \cite{afshordi2014emergent,stojkovic2013vanishing}. Thus, studying the phenomenon of dimensional reduction in short distances, a seemingly \textit{common thread} among the quantum gravity theories, has garnered significant interest in recent times.\par
In a discussion regarding dimensional reduction, the central element is the mechanism that enables such dimensional flow. In this paper, we consider the spacetime with a zero-point length as proposed in \cite{padmanabhan1985physical,padmanabhan1985planck,PhysRevLett.78.1854}. Having a minimum length or minimum resolvable distance in the spacetime of the order of Planck length ($\ell_{\rm Pl}=\sqrt{\hbar G/c^3}$) is also another ubiquitous feature among many different approaches towards quantum gravity \cite{hossenfelder2013minimal}. When the known principles of quantum physics and gravitation are applied simultaneously, Planck length appears as a fundamental lower bound to all physical length scales \cite{PhysRev.135.B849,padmanabhan1987limitations,garay1995quantum}; hence it is expected to be a robust feature of yet to be discovered quantum theory of gravitation. Moreover, as the minimal length acts as a UV regulator in the quantum field theories at four dimensions, it is expected to arise as a natural outcome of a correct quantum gravity theory \cite{dewitt1964gravity}. In the particular model we shall be dealing with here, the zero-point length appears due to the quantum fluctuation of spacetime. The Euclidean geodesic distance (squared) $\hat{\sigma}^2(x,x^\prime)$ between two events $x$ and $x^\prime$, when averaged over the quantum fluctuations of the ground state of spacetime, gets shifted, $\langle\hat{\sigma}^2\rangle=\sigma^2+\ell_0^2$, giving rise to the zero-point length, which is named aptly so since the geodesic distance does not vanish in the coincidence limit. The zero-point length $\ell_0$ is considered to be proportional to the Planck length, $\ell_0\propto\ell_{\rm Pl}$. One can also bring forth such a zero-point length by considering a duality symmetry in the path integral of a massive relativistic particle \cite{PhysRevLett.78.1854,PhysRevD.57.6206,PhysRevD.58.044009,shankaranarayanan2001hypothesis}, an idea which has been linked to T-duality in String theory \cite{fontanini2006zero,PhysRevLett.78.1854}. Moreover, it has been argued recently that in a certain limit, this approach also produces a deformed Heisenberg algebra \cite{mondal2020duality}, a natural starting point for many plausible phenomenological studies on quantum gravity. These connections motivate us to consider this approach further in its consequences on the ultraviolet dimensional reduction.\par
Although the most studied quantum analog of classical dimension is the spectral dimension, this is by no means the only way to study the ultraviolet dimensional reduction. The spectral dimension is defined in the context of a fictitious diffusion process on a manifold \cite{carlip2017dimension,ambjorn2005spectral}. When the diffusion occurs for a long duration, the particles probe more considerable distances, and as over large distances the quantum effects are not expected to contribute sufficiently, the spectral dimension coincides with the conventional topological dimension. On the other hand, when the characteristic diffusion time is sufficiently small, due to the quantum \textit{fluctuations} or \textit{fuzziness}, some of the dimensions are screened or become inaccessible to the diffusing particles resulting in a reduced spectral dimension for the spacetime.\par
Another way to define dimension is to consider free bosonic fields or relativistic ideal gas at finite temperature \cite{nozari2015high,amelino2017thermal}. As the thermal processes depend on the available degrees of freedom, the dimension information can be extracted from the thermodynamic quantities like the free energy. Similarly to the case of spectral dimension, we shall see that at low temperatures when the quantum gravity effects are negligible, the thermodynamic dimension coincides with the traditional topological value, whereas in sufficiently high temperature, the degrees of freedom available to the particles reduce; thus the thermodynamic dimension returns a smaller value than its traditional infrared counterpart.\par
Furthermore, another way of extracting dimension from a physical process is to consider the interaction between two point-like classical sources placed $r$ distance apart in a free field theory \cite{arzano2017non}. The potential energy of such interactions is similar in form to the Yukawa potential. Then, we compare the short distance form of the QG (quantum gravity) corrected potential with the conventional Yukawa potential's short distance behavior at various dimensions. The dimension at which the conventional Yukawa potential's scaling matches with the QG corrected one's scaling is to be considered the ultraviolet dimension of spacetime with zero-point length. We shall see that in the case of spacetime with (infrared) dimension four, the QG corrected potential assumes a constant value, which matches with the conventional Yukawa potential's behavior at dimensions $2$ for massive force carriers and at $3$ for massless force carriers. This allows us to interpret that the ultraviolet dimension of spacetime with zero-point length appears to be either three or two, depending on the nature of the force carriers.\par
Previously, the dimensional reduction for a spacetime with zero-point length at Planck scale has already been shown in \cite{2016}, by observing how the volume of a geodesic ball changes in that limit. There, the result is that the spacetime with zero-point length, at the Planck scale, appears to be of dimensions two, which is a typical value for the ultraviolet dimension in many different quantum theories of gravity. In this paper, we shall study the ultraviolet dimensional reduction in different contexts. In what follows, we shall study the spectral, thermodynamic, and interaction dimension of spacetime with a zero-point length. The study of ultraviolet dimensional reduction is vital because this phenomenon is supposed to be one among a handful of (possibly) observable predictions from quantum gravity \cite{mureika2011detecting,anchordoqui2012vanishing,amelino2013dimensional}.
\section{Spectral dimension}\label{spectral}
Consider a diffusion process in a $D$ dimensional Euclidean space governed by the heat equation
\begin{equation}\label{eq:heat_equation}
    \left[\frac{\partial}{\partial s}-\Delta_x\right]\mathcal{K}(x,x^\prime;s)=0,
\end{equation}
where $s$ is an auxiliary time parameter with dimension of length squared, $\Delta_x$ is the Laplacian associated with the geometry of the Euclidean space and $\mathcal{K}(x,x^\prime;s)$ is the heat kernel quantifying the probability density of a phenomenon that a particle starts off from an event $x$ at $s=0$ and then diffuses to another event $x^\prime$ in diffusion time $s$. The heat kernel, additionally, satisfies the boundary condition
\begin{align}
    \mathcal{K}(x,x^\prime;0)=\frac{\delta^{(D)}(x-x^\prime)}{\sqrt{g(x)}},
\end{align}
where, $\delta^{(D)}(x-x^\prime)$ is a $D$ dimensional Dirac delta distribution and $g$ is the determinant of the Euclidean metric $g_{ab}$ of the manifold.\par
It is important to note that although $s$ is a ``fake time,'' it nevertheless provides the scale at which the diffusion process is being probed. Small value of $s$, that is, $s\to 0$ corresponds to the probing of ultraviolet scale, whereas large values, i.e., $s\to\infty$ corresponds to the probing of infrared scale.\par
An useful quantity can be deifined called the return probability, which quantifies the probability that a particle starting from $x$ in the diffusion process returns to the same location after a diffusion time $s$ has passed
\begin{equation}
    P(s)\equiv{\rm Tr}\mathcal{K}(x,x;s)=\frac{\int {\rm d}^Dx\,\sqrt{g(x)}\,\mathcal{K}(x,x;s)}{\int {\rm d}^Dx\,\sqrt{g(x)}}.
\end{equation}
Then the spectral dimension is defined in terms of the return probability as
\begin{equation}\label{eq:spectral dimension}
    \mathbb{D}_S(s)=-2\frac{{\rm d}\log P(s)}{{\rm d}\log s}.
\end{equation}
The motivation behind this definition stems from the fact that in flat Euclidean space the return probability $P(s)$ has the form $(4\pi s)^{-\frac{D}{2}}$, and thus the spectral dimension coincides with the conventional topological dimension of the manifold, i.e., $\mathbb{D}_S(s)=D$.\par
In principle, as the Laplacian contains information regarding the curvature of the manifold, through the solution $\mathcal{K}(x,x^\prime;s)$ and then through $P(s)$, that information is passed onto the spectral dimension itself making it a curvature or geometry dependent quantity. However, as we shall be studying the quantum gravity effects in a mesoscopic region near the Planck scale, hopefully, much below the radius of curvature of the given manifold, curvature effects can be safely neglected and henceforth, we can get away by working with the flat Euclidean metric.\par
The heat kernel for the spacetime with zero-point length $\mathcal{K}_{\ell_0}(x,x^\prime;s)$ is already known and it reads in terms of the conventional heat kernel solution as
\begin{align}\label{eq:heat kernel}
    \mathcal{K}_{\ell_0}(x,x^\prime;s)&=e^{-\frac{\ell_0^2}{4s}}\mathcal{K}_{\rm std.}(x,x^\prime;s)\nonumber\\
    &={(4\pi s)^{-\frac{D}{2}}}\exp\left\{-\frac{(x-x^\prime)^2+\ell_0^2}{4s}\right\},
\end{align}
which can be obtained, among other methods, simply by replacing the geodesic interval $\sigma^2$ by $\sigma^2+\ell_0^2$ \cite{PhysRevLett.78.1854,PhysRevD.57.6206,PhysRevD.58.044009}. This additional QG correction factor exponentially suppresses the probability of diffusion for time scales below the zero-point length ($\ell_0^2/4\gg s$), that is, physical processes occurring below the Planck scale is suppressed.\par
Given this, it is straightforward to calculate the spectral dimension
\begin{align}
    \mathbb{D}_{S,\ell_0}(s)=D-\frac{\ell_0^2}{2s}.
\end{align}
To understand this result, let us first consider the infrared case, $s\to\infty$. In this limit, the spectral dimension coincides with the conventional topological dimension of the spacetime, reinforcing the fact that at large length scales, the quantum gravity effects are weak, and the diffusion processes occur as if they were on a smooth manifold. For small values of $s>\ell_0^2$, the spectral dimension starts to deviate from the topological dimension $\mathbb{D}_{S,\ell_0}<D$ as the quantum fluctuations or the \textit{fuzziness} or \textit{blurriness} of spacetime starts interfering with the diffusion processes preventing those from accessing all the dimensions. When the diffusion processes occur at the characteristic diffusion time of the size of the zero-point length, the spectral dimension flows from $D$ to $D-\frac{1}{2}$. In four dimensions, then, the Planckian spectral dimension of spacetime with zero-point length is 3.5. This non-integer nature of the spectral dimension ($D-\frac{1}{2}$) is the testament of its fractal nature in the ultraviolet region.\par
Moreover, when the diffusion time becomes one-fourth of the zero-point length, the spacetime dimension reduces to $D-2$ (or 2 for a (topologically) four-dimensional spacetime). Below the value 2, as the spectral dimension keeps on decreasing towards $-\infty$ as $s\to 0$. The negativity of the dimension can be interpreted as follows. In \cite{nicolini2011hausdorff}, the authors realize that, for fractals, negative dimension is a measure of the degree of the emptiness of empty sets \cite{mandelbrot1990negative}, and hence, the negativity of Hausdorff dimension of quantum mechanical paths imply that in the \textit{trans-Planckian regime} the quantum paths of the particles gets entirely disintegrated by the high degree of fluctuations in the underlying (quantum) manifold. Moreover, the diffusion heat equation (1) is just a Wick rotated Schr\"odinger equation for a non-relativistic particle, and hence, the Hausdorff dimension and the spectral dimension are interrelated. As a path is the history of a particle through spacetime events, its dissolution or disintegration can be perceived as local emptiness in the manifold, as the particle finds it difficult to reach the next spacetime event of its history owing to the huge fluctuations of spacetime, thereby the notion of negative dimension associated with empty sets might be applicable.\par
The dimensional reduction also relates to reducing the degrees of freedom or dimensions available to the diffusion process. As the QG correction due to quantum fluctuations of spacetime suppresses processes that occur below a particular length scale, near that scale, the suppression manifests itself in a reduction of available degrees of freedom. Therefore, the dimension appears to be reduced. The exponential nature of the suppression seems to be the reason behind a diverging spectral dimension in the limit $s\to 0$.\par
The divergence of spectral dimension in the deep ultraviolet limit might immediately seem off-putting; however, such divergences have appeared in other quantum spacetime models as well \cite{PhysRevD.102.086003,mandrysz2018ultralocal}.
\section{Thermodynamic dimension}\label{thermodynamic}
The definition of thermodynamic dimension can be motivated from the observation that free energy per unit volume of space, $F(\beta)/\mathcal{V}_d$, of a bosonic free field or ideal gas at a finite temperature $T=\frac{1}{\beta}$ in a $D=d+1$ dimensional spacetime scales as $\beta^{-D}$. Thus, in a similar vein to (\ref{eq:spectral dimension}) we define the thermodynamic dimension of a system as follows
\begin{equation}
    \mathbb{D}_{T}=-\frac{{\rm d}\log \left(\frac{F(\beta)}{\mathcal{V}_d}\right)}{{\rm d}\log \beta},
\end{equation}
here, $\mathcal{V}_d$ is the volume of the finite region, \textit{the box}, over which we are averaging the free energy. This definition is equivalent to the observation in \cite{carlip2017dimension,atick1988hagedorn} that in quantum field theory in $d$ dimensions, $F/VT$ always grows at least as $T^{D-1}$, from which the dimension of the spacetime can be deduced.\par
As we already know the heat kernel solution (\ref{eq:heat kernel}) for spacetime with a zero-point length, it is pragmatic to apply Schwinger's proper time formalism to study the finite temperature case as well. A reader unfamiliar with this formalism might consult \cite{das2019field} for an introduction on the subject.\par
To transition to a case with finite temperature, we start with the standard approach of compactifying the Euclidean time dimension into a closed manifold $\mathbb{S}^1$ with an orbit length or period equal to the inverse temperature $\beta$. Thus the $D$ dimensional flat Euclidean spacetime now has the structure $\mathbb{R}^d\times\mathbb{S}^1$ and the periodicity in the Euclidean time brings forth a finite temperature. In such a space, we shall compute the trace of the finite temperature heat kernel, which can be related to the thermal partition function of the system.\par 
The thermal heat kernel satisfies a periodicity condition $\mathcal{K}^{\beta}(\tau,\mathbf{x};\tau^\prime+\beta,\mathbf{x}^\prime|s)=\mathcal{K}^{\beta}(\tau,\mathbf{x};\tau^\prime,\mathbf{x}^\prime|s)$, here, $\tau$ is the Euclidean time and $\mathbf{x}$ are the coordinates on $\mathbb{R}^d$. To encompass this periodicity one sums over the standard heat kernel $\mathcal{K}(\tau,\mathbf{x};\tau^\prime+n\beta,\mathbf{x}|s)$ for all possible values of $n$, that is, one gets the thermal heat kernel from the standard one by doing the following sum
\begin{align}
    \mathcal{K}^{\beta}(\tau,\mathbf{x};\tau^\prime,\mathbf{x}^\prime|s)=\sum_{n=-\infty}^\infty\mathcal{K}_{\rm std.}(\tau,\mathbf{x};\tau^\prime+n\beta,\mathbf{x}^\prime|s).
\end{align}
To compute the trace of the thermal heat kernel, we have to identify the event $(\tau,\mathbf{x})$ and $(\tau^\prime,\mathbf{x}^\prime)$ and integrate over the volume of the box. Now, consider a geodesic of a particle starting off from $x=(\tau,\mathbf{x})$ and propagating to $x=(\tau+\beta,\mathbf{x}^\prime)$. If we identify $\mathbf{x}$ and $\mathbf{x}^\prime$, we get a loop which is incontractible in the given geometry $\mathbb{R}^d\times\mathbb{S}^1$. Thus the trace corresponds to summing over all possible loops of this kind. This construction has been proposed in \cite{atick1988hagedorn} (also see, \cite{gusev2015finite,shevchenko2016archimedes}). After incorporating the zero-point length correction from quantum gravity, the trace of the thermal heat kernel reads as follows
\begin{align}
    {\rm Tr}\mathcal{K}^\beta_{\ell_0}(s)&=\int{\rm d}^{d+1}x \,\, e^{-\frac{\ell_0^2}{4s}}\sum_{n=-\infty}^{\infty}\mathcal{K}_{{\rm std.}}(\tau,\mathbf{x};\tau+n\beta,\mathbf{x}|s)\nonumber\\
    &=e^{-\frac{\ell_0^2}{4s}}\int_0^\beta {\rm d}\tau \sum_{n=-\infty}^{\infty}\mathcal{K}^{\,(\tau)}(\tau,\tau+n\beta;s)\nonumber\\
    &\qquad\times\int_{\rm Box}{\rm d}^{d}x \,\, \mathcal{K}(\mathbf{x},\mathbf{x};s)\nonumber\\
    &=e^{-\frac{\ell_0^2}{4s}}\left(\frac{\beta}{(4\pi s)}\sum_{n=-\infty}^{\infty}e^{-\frac{\beta^2n^2}{4s}}\right)\left(\frac{\mathcal{V}_d}{(4\pi s)^{\frac{d}{2}}}\right)\nonumber\\
    &=\frac{\beta\mathcal{V}_d}{(4\pi s)^{\frac{d+1}{2}}}\sum_{n=-\infty}^{\infty} e^{-\frac{\beta^2 n^2+\ell_0^2}{4s}}.
\end{align}
In principle, due to the finite nature of the box, there should be a term proportional to the surface area of the box $\mathcal{S}_{d-1}$. However, as we are interested in the result of the form ${F(\beta)}/{\mathcal{V}_d}$, the surface area term being proportional to ${\mathcal{S}_{d-1}}/{\mathcal{V}_d}$ becomes sub-leading in the large box approximation. Moreover, as the surface term in the expression for free energy scales as $\beta^{D-1}$, at sufficiently high temperature, this can be neglected in favor of the volume-dependent term. Given the trace, we define the (logarithm of) partition function as
\begin{align}
    \log Z_{\ell_0}(\beta)=\frac{1}{2}\int_0^{\infty}\frac{{\rm d}s}{s}{\rm Tr}\mathcal{K}^\beta_{\ell_0}(s).
\end{align}
In the conventional field theory at finite temperature for a massless free scalar field, this integral diverges for the term corresponding to $n=0$ as the integration over the Matsubara zero-mode diverges due to an ill-defined Mellin transform \cite{strickland2019relativistic}\footnote{Also see, \textit{Thermal Quantum Field Theory} by Gergely Endr\H odi, (2018) \url{https://itp.uni-frankfurt.de/\~endrodi/thermalFT.pdf}}. However, in our case, the zero-point length $\ell_0$ acts as a regulator just as a mass parameter would do to make the term finite.\par
The free energy of a thermal system is related to the partition function via the following relation
\begin{equation}
    F_{\ell_0}(\beta)=-\frac{1}{\beta}\log Z_{\ell_0}(\beta).
\end{equation}
Finally, we can express the QG corrected free energy of a spacetime with zero-point length as follows
\begin{align}
    F_{\ell_0}(\beta)=&-\mathcal{V}_d\int_0^\infty \frac{{\rm d}s}{s}{(4\pi s)^{-\frac{D}{2}}}\sum_{n=1}^{\infty}e^{-\frac{\beta^2 n^2+\ell_0^2}{4s}}-\frac{\mathcal{V}_d}{2}\frac{\Gamma\left(\frac{D}{2}\right)}{\ell_0^D\pi^{\frac{D}{2}}}.
\end{align}
In the limit $\ell_0\to 0$, the first term in the above expression reduces to the standard expression in statistical thermodynamics, whereas the second term diverges. As the zero-point length acts as a regulator in lieu of mass, taking the limit $\ell_0\to 0$ corresponds to taking the limit $m\to 0$ in standard field theory, in which the massless case has a divergent part. For our purpose, most of the time, the second term can be ignored since it is independent of $\beta$ and contains no information regarding dimension. We, therefore, focus on computing the first term.\par
To compute the sum and integral in the first term of the above expression, we can define a new variable $y=\frac{\beta^2}{4s}$ and rewrite the expression for free energy 
\begin{align}\label{eq: free energy 1}
    F_{\alpha}(\beta)=-\frac{\mathcal{V}_d}{\beta^D \pi^{D/2}}\sum_{n=1}^{\infty}\int_0^\infty {\rm d}y \, y^{\frac{D}{2}-1}e^{-y(n^2+\alpha^2)},
\end{align}
where $\alpha=\frac{\ell_0}{\beta}$. This new parameter $\alpha$ will set the scale at which the system operates. At very low temperature ($\beta\to\infty$) the parameter $\alpha$ vanishes restoring the conventional physics and the effect of the zero-point is not felt, whereas at very high temperature ($\beta\to 0$) the parameter assumes a high value $\alpha\gg1$, the limit of our interest. The parameter value $\alpha\sim\mathcal{O}(1)$, corresponds to the Planck energy scale.\par
After integration, (\ref{eq: free energy 1}) reduces to (see appendix \ref{appendix A})
\begin{align}\label{eq:Gamma-type integration}
    F_{\alpha}(\beta)=-\frac{\mathcal{V}_d}{\beta^D \pi^{D/2}}\Gamma\left(\frac{D}{2}\right)\sum_{n=1}^{\infty}\frac{1}{(n^2+\alpha^2)^{\frac{D}{2}}}.
\end{align}
The infinite sum above is a generalized Mathieu series and can be expressed in the following integral form, which would be more helpful for us in considering the various limits of the parameter $\alpha$ (see Appendix \ref{Appendix B})
\begin{align}
    F_{\alpha}(\beta)&=-\frac{\mathcal{V}_d}{\beta^D }\left({2\alpha}{\pi}\right)^{-\nu}\int_{0}^{\infty}\frac{x^{\nu}J_{\nu}(\alpha x)}{e^x-1}{\rm d}x\nonumber\\
    &=-\mathcal{V}_d\beta^{-D}\left({2}{\pi}\right)^{-\nu} \alpha^{-\nu}\mathcal{F}_\nu(\alpha),
\end{align}
where $\nu=\frac{D}{2}-\frac{1}{2}$. As we shall be always considering $D>1$, $\nu$ should be always greater than or equal to $\frac{1}{2}$. There are two $\alpha$ dependent components in the above expression for the free energy, one, the $\alpha^{-\nu}$ part and two, the Hankel transorm
\begin{align}
    \mathcal{F}_\nu(\alpha)=\int_{0}^{\infty}\frac{x^{\nu}J_{\nu}(\alpha x)}{e^x-1}{\rm d}x
\end{align}
part. The combination of the behaviour of these two components in various limits determines the overall $\alpha$ dependency of the free energy. Now we shall examine case by case the asymptotic forms of the above expression in the various limits of our interest.
\subsection{Low temperature (\texorpdfstring{$\alpha\ll 1$}{all1}) limit}
For $\alpha\ll1$ the Bessel function has an asymptotic form
\begin{align}
    J_\nu(\alpha x)\simeq\frac{(x\alpha)^{\nu}}{2^\nu \Gamma(\nu+1)}.
\end{align}
Putting this into the Hankel transform returns
\begin{align}
    \mathcal{F}_{\nu}(\alpha)\simeq\frac{\left(2{\alpha}\right)^\nu}{\sqrt{\pi}}\zeta(2\nu+1)\Gamma\left(\nu+\frac{1}{2}\right)
\end{align}
in the limit $\alpha\to 0$. The Hankel transforms scales with $\alpha^\nu$ which is canceled by the $\alpha^{-\nu}$ factor multiplying it in the expression for the free energy, thus for small $\alpha$ we have
\begin{align}
    F_{\alpha}(\beta)=-\frac{\mathcal{V}_d}{\pi^{\frac{D}{2}}} \beta^{-D}\Gamma\left(\frac{D}{2}\right).
\end{align}
This is exactly the result one obtains if $\alpha$ is set to zero in (\ref{eq:Gamma-type integration}). Recall that we are ignoring the $n=0$ term and focusing only on the part which directly corresponds to the similar result from statistical thermodynamics.\par
In this limit the thermal dimension agrees with the topological dimension
\begin{align}
    \mathbb{D}_T=-\frac{{\rm d}\log\left(\frac{F_{\alpha}}{\mathcal{V}_d}\right)}{{\rm d}\log\beta}=D.
\end{align}
The equation of state parameter in this limit
\begin{align}
    w=-\frac{\frac{\partial}{\partial \mathcal{V}_d}\left({F_{\alpha}(\beta)}\right)}{\frac{\partial}{\partial\beta}\left(\beta\frac{F_{\alpha}(\beta)}{\mathcal{V}_d}\right)}=\frac{1}{D-1},
\end{align}
which again agrees with the conventional results in statistical thermodynamics. For a(n) (infrared) $D=4$ dimensional spacetime the equation of state parameter is $w={1}/{3}$, and effectively the massless bosonic free field or the relativistic particles behaves like radiation.\par
As we have neglected the $n=0$ term our results resembles to that obtained in the statistical thermodynamics which in turn corresponds to the high temperature ($\beta\ll1$) limiting expressions in the conventional field theory. Thus our results here corresponds to sufficiently high energy scales to have the $n=0$ term ignored but still much below Planckian level ($\alpha<1$) to catch any quantum gravity effects.
\subsection{High temperature (\texorpdfstring{$\alpha\gg 1$}{agg1}) limit}
At very high temperatures, much beyond the Planck level itself, deep into the ultraviolet regime ($\alpha\gg1$) the Hankel transform has the following asymptotic form (see, Appendix \ref{Appendix C})
\begin{align}
    \mathcal{F}_{\nu}(\alpha)\simeq \frac{\Gamma(\nu)}{2}\left(\frac{2}{\alpha}\right)^{\nu}-\frac{\Gamma(\nu+\frac{1}{2})}{4\sqrt{\pi}}\left(\frac{2}{\alpha}\right)^{1+\nu}.
\end{align}
Therefore, in this limit, the expression for the free energy $F_{\alpha}(\beta)$ reads
\begin{align}
    -\mathcal{V}_d\beta^{-D}\left[\frac{\Gamma\left(\frac{D-1}{2}\right)}{2\pi^{\frac{D-1}{2}}}\left(\frac{\beta}{\ell_0}\right)^{D-1}-\frac{\Gamma\left(\frac{D}{2}\right)}{2\pi^{\frac{D}{2}}}\left(\frac{\beta}{\ell_0}\right)^{D}\right].
\end{align}
We note that the second term in the asymptotic expression gets precisely canceled by the $n=0$ term, $-\frac{\mathcal{V}_d}{2}\frac{\Gamma\left(\frac{D}{2}\right)}{\ell_0^D\pi^{\frac{D}{2}}}$, which we had ignored so far. Finally, the free energy reduces to
\begin{align}
    F_{\alpha}(\beta)\simeq -\mathcal{V}_d\frac{\Gamma\left(\frac{D-1}{2}\right)}{2\pi^{\frac{D-1}{2}}}\left(\frac{\beta^{-1}}{\ell_0^{D-1}}\right).
\end{align}

Therefore, the thermodynamic dimension of the spacetime with zero-point length in the deep ultraviolet regime is
\begin{align}
    \mathbb{D}_T=-\frac{{\rm d}\log\left(\frac{F_{\alpha}}{\mathcal{V}_d}\right)}{{\rm d}\log\beta}=1.
\end{align}
The ultraviolet thermodynamic dimension does not depend on the infrared dimension $D$. The spacetime fusing to have dimension one is very difficult to interpret with any traditional intuition. Moreover, in this limit, much like the spectral dimension, the density of state parameter is not well defined, which again reinforces the fact that this approximate method of adding a zero-point length to the geodesic interval works only in the mesoscopic region near the Planck scale and at best up to the Planck scale itself. However, beyond the Planck scale, this approach might not be reliable and has to be replaced with a more sophisticated complete theory of quantum gravity.
\subsection{Near the Planck scale (\texorpdfstring{$\alpha\sim\mathcal{O}(1)$}{aO1})}
\begin{figure}[ht]
    \centering
    \includegraphics[width=0.48\textwidth]{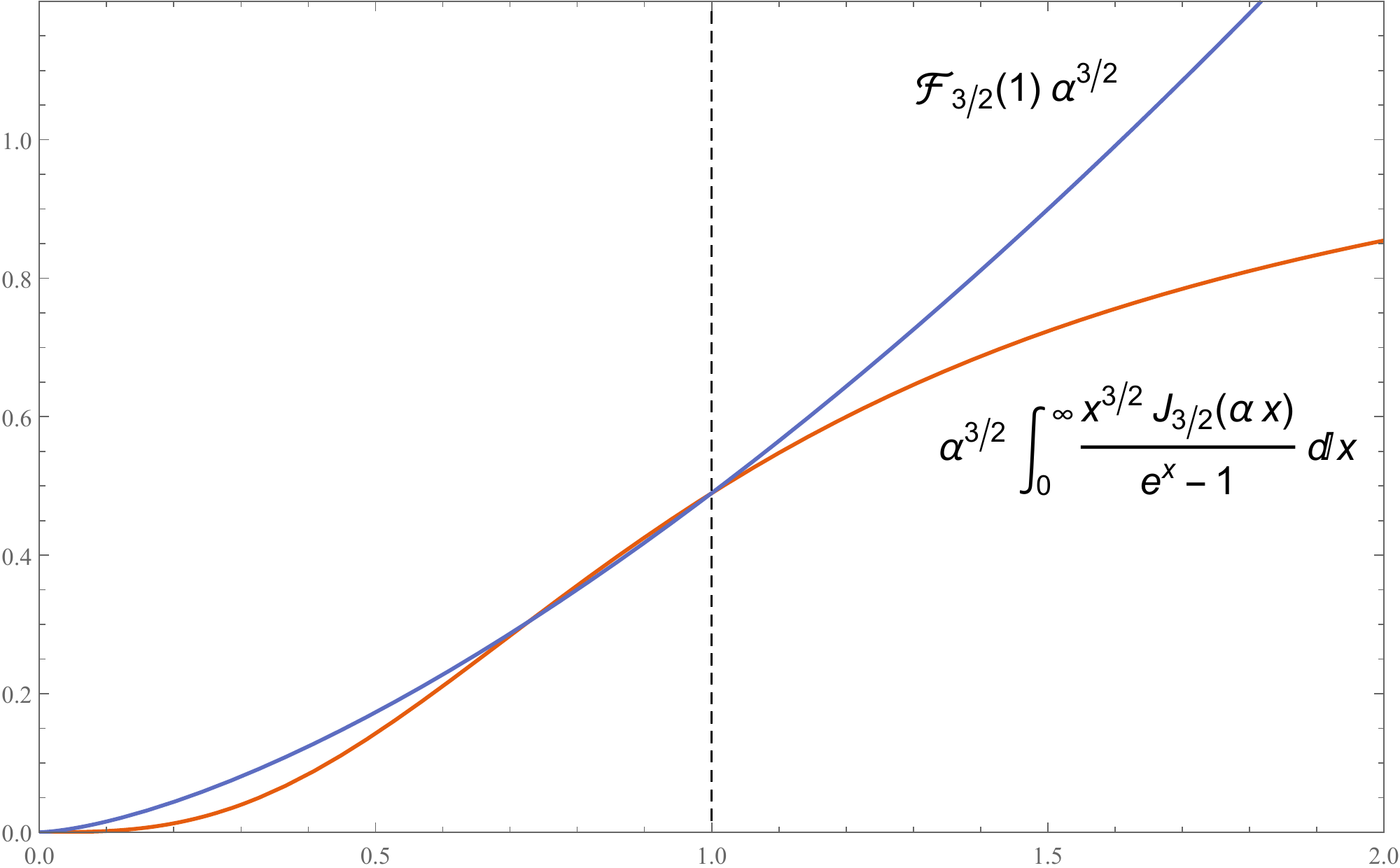}
    \caption{A numerical demonstration in $D=4$ to the fact that near $\alpha\sim1$ the free energy can be approximated by a curve of the form $\alpha^{\nu}\mathcal{F}_{\nu}(1)$.}
    \label{fig:my_label}
\end{figure}

It is trickier to deal with the case when $\alpha\to 1$ because none of the asymptotic forms shown above works in this limit. Now, we see that the Hankel transform $\mathcal{F}_{\nu}(\alpha)$ has a maximum somewhere around $\alpha\sim1$ because in the region $\alpha< 1$, its asymptotic dependency on $\alpha$ is of the form $\alpha^p$ with $p>0$, that is, the function is monotonically increasing in this region. However, on the other hand, in the region $\alpha>1$, its asymptotic dependency is of the form $\alpha^{q}$ with $q<0$; thus, the function monotonically decreases. Therefore, somewhere near $\alpha\sim1$, $\mathcal{F}_{\nu}(\alpha)$ reaches its maxima, i.e., its asymptotic dependency takes a form $\alpha^m$ such that $m\to 0$ as $\alpha\to 1$. Therefore, for the free energy
\begin{align}
    F_{\alpha}(\beta)&=-\mathcal{V}_d\ell_0^{-D}\left({2}{\pi}\right)^{-\nu} \alpha^{D-\nu}\mathcal{F}_\nu(\alpha),
\end{align}
its $\alpha$ dependence near $\alpha\sim\mathcal{O}(1)$ can be approximated by the $\alpha$ dependent factor multiplying the Hankel transform as follows
\begin{align}
    F_{\alpha}(\beta)&\sim-\mathcal{V}_d\ell_0^{-D}\left({2}{\pi}\right)^{-\nu} \alpha^{\nu}\mathcal{F}_\nu(\alpha=1),
\end{align}
where, we have used the fact that $D-\nu=\nu$. In the case of $D=4$, we demonstrate this fact below in a numerical plot (figure \ref{fig:my_label}).\par
Here we have plotted the free energy (modulo some irrelevant prefactors) against $\alpha$, and we observe that the curve $\alpha^{\nu}\mathcal{F}_{\nu}(1)$ is a good approximation to the free energy curve near the point $\alpha\sim1$. However, in the region $\alpha>1$ or $\alpha<1$ these two curves deviate significantly from each other as expected.\par
Therefore, as we approach the Planckian energy scale ($\alpha\to 1$), the thermodynamic dimension approaches the value
\begin{align}
    \lim_{\alpha\to 1}\mathbb{D}_{T,\alpha}=\nu=\frac{D-1}{2}.
\end{align}
For a spacetime with infrared dimension $D=4$, at Planck scale, the thermodynamic dimension reduces to the value $1.5$. Again, like the spectral dimension, the thermodynamic dimension as well has a non-integer value near the Planck scale.\par
The equation of state parameter near the Planck scale
\begin{align}
    w_{\rm Planck}=\frac{1}{\nu-1}=\frac{2}{D-3}.
\end{align}
For $D=4$, near the Planck scale the equation of state parameter has the value $w_{\rm Planck}=2$. Therefore, due to the presence of the zero-point length in the quantum spacetime, a bosonic ideal gas which behaved like radiation at very small energy scales undergoes a continuous phase transition and near the Planck scale behaves as if it is a fluid with equation of state parameter 2. Thus at the high energy the existence of quantum fluctuations transforms radiation into a `stiffer' fluid.
\section{Dimension from two point source interaction}\label{interaction}
In a free field theory, the interaction potential $V_{D}(r)$ between two static classical sources placed at $r$ distance apart in $D$ dimensions has a form similar to the Yukawa potential. Thus, if we can calculate $V_{D}(r)$ in the case of a zero-point length spacetime, we shall be able to extract out the dimension information of spacetime at microscopic length scales by considering the asymptotic expression of $V_{D}(r)$ in the limit $r\to 0$ and then by comparing it to the scaling behavior of the Yukawa potential in that limit in different dimensions. Say, there is a dimension $\mathcal{D}$ in which the short distance scaling behavior of the Yukawa potential matches with the ultraviolet limit of the $D$ dimensional QG corrected potential. Then this observation can be said to have the interpretation that for spacetime with zero-point length, the infrared dimension $D$ flows to a different value $\mathcal{D}$ in the ultraviolet limit. Let us make this assertion explicit in the following.\par
To not make things mathematically too complicated, we shall calculate the zero-point length corrected potential in a spacetime with infrared dimension four and then examine how this number changes in the ultraviolet limit. In a spacetime with zero-point length, the kinetic structure of a free scalar field or the d'Alembertian operator is supposed to change from what we use in the conventional case \cite{padmanabhan2020class}. Then, the action of a free scalar field can be expressed in the following form
\begin{align}
    A&\equiv-\frac{1}{2}\int {\rm d}^4 x\, \phi(x)F\left[\left(-\Box\right)\right]\phi(x)\nonumber\\
    &=-\frac{1}{2}\int \frac{{\rm d}^4 p}{(2\pi)^4}\, \phi^*(p)F\left[\left(p^2\right)\right]\phi(p),
\end{align}
where, we have modified the d'Alembertian operator $-\Box$ to a new form $F\left[\left(-\Box\right)\right]$. Then the generating functional for this particular action simply reads as follows,
\begin{align}
    Z[J]=\int\mathcal{D}\phi(x)\, \exp\left[iA\left[\phi(x)\right]+i\int {\rm d}^4 x\, J(x)\phi(x)\right].
\end{align}
For a free field theory, the path integral can be performed explicitly and can be shown \cite{zee2010quantum} to have the form
\begin{align}
    Z(J)=Z(J=0)e^{iW(J)},
\end{align}
where, $W(J)$ is
\begin{align}\label{eq:energy}
    W(J)=-\frac{1}{2}\int {\rm d}^4x {\rm d}^4y J(x)D(x-y) J(y),
\end{align}
with $D(x-y)$ being the Green's function for the modified d'Alembertian operator satisfying the equation
\begin{align}
    F\left[\left(-\Box\right)\right]D(x-y)=\delta^{(4)}(x-y).
\end{align}
This equation can be solved using the usual method of utilizing Fourier transform and the Green's function can then be expressed in the Fourier space as
\begin{align}
    D(x-y)=\int \frac{{\rm d}^4 p}{(2\pi)^4}\frac{e^{ip\cdot (x-y)}}{F\left[\left(p^2\right)\right]+i\epsilon},
\end{align}
here, $+i\epsilon$ is the standard Feynman regulator. Now, we go back to our case of an interaction between two static classical sources. These sources can be modeled with three dimensional Dirac delta distribution. We take $J$ as a superposition of two delta distribution sources
\begin{align}
    J(x)=\sum_{a=1}^2 J_a(x); \qquad J_a(x)=\delta^{(3)}(\mathbf{x}-\mathbf{x}_a).
\end{align}
Plugging this in (\ref{eq:energy}) and ignoring self interaction terms we can write
\begin{align}
    W(J)&=-\int\int {\rm d}x^0 {\rm d}y^0\int\frac{{\rm d}p^0}{2\pi}e^{ip^0(x-y)^0}\nonumber\\
    &\hspace{1cm}\times\int \frac{{\rm d}^3 \mathbf{p}}{(2\pi)^3}\frac{e^{i\mathbf{p}\cdot(\mathbf{x}_1-\mathbf{x}_2)}}{F\left[\left(p^2\right)\right]+i\epsilon}.
\end{align}
Integrating over $y^0$ produces a delta function and sets $p^0$ to zero. Therefore, $W(J)$ reduces to
\begin{align}
    W(J)=-\left(\int{\rm d}x^0\right)\int \frac{{\rm d}^3 \mathbf{p}}{(2\pi)^3}\frac{e^{i\mathbf{p}\cdot\mathbf{r}}}{F\left[\left(\mathbf{p}^2\right)\right]},
\end{align}
here the separation vector is denoted as $\mathbf{r}=\mathbf{x}_1-\mathbf{x}_2$. The $+i\epsilon$ has been dropped assuming $F\left[\left(\mathbf{p}^2\right)\right]$ is always a positive quantity. Realizing that the path integral $Z(J)=Z(0)e^{iW(J)}$ corresponds to $\langle 0|e^{-i\hat{H}T}|0\rangle=e^{-iET}$, where $E$ is the energy due to the interaction between the sources, and considering $\int {\rm d}x^0$ to be the interaction time, one extracts out the energy of the configuration of the classical sources, i.e., the potential energy
\begin{align}\label{eq:potential_integration}
    V^{\ell_0}(r)&=-\int \frac{{\rm d}^3 \mathbf{p}}{(2\pi)^3}\frac{e^{i\mathbf{p}\cdot\mathbf{r}}}{F\left[\left(\mathbf{p}^2\right)\right]}\nonumber\\
    &=-\frac{1}{2\pi^2r}\int_0^\infty{\rm d}\left(|\mathbf{p}|\right) \,\,|\mathbf{p}|\frac{\sin(|\mathbf{p}|r)}{F\left[\left(|\mathbf{p}|^2\right)\right]}.
\end{align}
So far we have not taken any particular form for the modified d'Alembertian operator $F\left[\left(-\Box\right)\right]$. For spacetime with zero-point length, changing the geodesic distance to the following form $\sigma^2\to\sigma^2+\ell_0^2$ modifies the d'Alembertian in the follwoing form \cite{padmanabhan2020class}
\begin{align}
    F\left[\left(\mathbf{p}^2\right)\right]=\frac{\sqrt{|\mathbf{p}|^2+m^2}}{\ell_0 K_{1}\left(\ell_0\sqrt{|\mathbf{p}|^2+m^2}\right)}.
\end{align}
For other studies on similar UV modified propagator or Green's function, see \cite{wondrak2019constraints,kan2021vacuum,kan2021discrete}.\par 
Now, using the formula (6.726.3) in \cite{gradshte2007table}, we can perform the integration
\begin{align}
    &\int_0^\infty{\rm d}\left(|\mathbf{p}|\right) \,\,\frac{|\mathbf{p}|}{\sqrt{|\mathbf{p}|^2+m^2}}{\sin(|\mathbf{p}|r)}\,\ell_0 K_1(\ell_0\sqrt{|\mathbf{p}|^2+m^2})\nonumber\\
    &=\sqrt{\frac{m\pi}{2}}r\left(r^2+\ell_0^2\right)^{-\frac{1}{4}}K_{-\frac{1}{2}}\left(m\sqrt{r^2+\ell_0^2}\right).
\end{align}
Therefore, the potential energy in this case has the form
\begin{align}
    V^{\ell_0}(r)=-\frac{\sqrt{m}}{(2\pi)^{\frac{3}{2}}}\left(r^2+\ell_0^2\right)^{-\frac{1}{4}}K_{-\frac{1}{2}}\left(m\sqrt{r^2+\ell_0^2}\right).
\end{align}
In the limit $r\gg\ell_0$, the potential energy presumes its usual Yukawa potential form
\begin{align}
    V^{\ell_0}(r)=-\frac{1}{4\pi r}e^{-mr},
\end{align}
whereas, in deep ultraviolet region ($r\to 0$) reduces to a constant
\begin{align}
    V^{\ell_0}(r)=-\frac{e^{-m\ell_0}}{4\pi \ell_0}.
\end{align}

We must compare these results to the Yukawa-like potential in an arbitrary dimension derived in the conventional QFT without QG correction. A general expression for the Yukawa-like potential in the $D=d+1$ spacetime dimensions has been derived in appendix \ref{Appendix D}. Explicitly, the potential has the following expression
\begin{align}
    V_{d+1,\,\rm Yukawa}(r)=-\frac{1}{(2\pi)^{\frac{d}{2}}}\left(\frac{m}{r}\right)^{\frac{d-2}{2}}K_{\frac{d-2}{2}}(mr).
\end{align}
Let us see its asymptotic scaling behaviors in both short and large distance scales in different dimensions
\begin{align}
    \begin{array}{ccc}
    {\rm Dimension} & {\rm Ultraviolet} & {\rm Infrared} \\
    D=2 &  \sim{\rm constant} & \sim -e^{-mr},\\
    D=3 & \sim \log r & \sim -{e^{-mr}}/{\sqrt{r}},\\
    D=4 & \sim -{1}/{r} & \sim -{e^{-mr}}/{r}, \\
    D=5 & \sim -{1}/{r^2} &\sim -{e^{-mr}}/({r\sqrt{r}}).
    \end{array}
\end{align}
Therefore, we see that the zero-point length corrected potential's scaling behavior in large distances matches with that of Yukawa-like potential for $D=4$ as expected, whereas in short distances its behavior matches with that of Yukawa-like potential for $D=2$. Therefore, we conclude that while using massive force carriers to probe the nature of the spacetime with zero point length, it appears that its dimension flows from the expected infrared value $4$ to a reduced value $2$ in the deep ultraviolet regime.\par
Now, let us consider the case for massless force carries. Then the integration in (\ref{eq:potential_integration}) can be done using the formula (6.671.5) in \cite{gradshte2007table} and has the following form
\begin{align}
    \int_0^\infty{\rm d}\left(|\mathbf{p}|\right) \,\,{\sin(|\mathbf{p}|r)}\,\ell_0 K_1(\ell_0|\mathbf{p}|)=\frac{\pi r}{2\sqrt{r^2+\ell_0^2}}.
\end{align}
Thus the potential energy in this case reads
\begin{align}
    V(r)=-\frac{1}{4\pi\sqrt{r^2+\ell_0^2}}.
\end{align}
We see that in the limit $\ell_0\to 0$, the expression for potential energy reduces to its usual form $-1/(4\pi r)$. Now, in the deep ultraviolet region ($r\to 0$), whereas the conventional potential energy diverges, the QG corrected potential reaches a constant value
\begin{align}
    \lim_{r\to 0}V(r)=-\frac{1}{4\pi \ell_0}={\rm constant}.
\end{align}
Comparing that with Yukawa-like potential's scaling behavior for massless particles case, which is $V_{D,\,{\rm Yukawa,}\,m=0}(r)\sim r^{3-D}$ (appendix \ref{Appendix D}) implies at short distances $D=3$. Therefore, due to the quantum gravity effects at small scales, the dimension of spacetime has reduced to a value 3, whereas in the infrared regime ($r\gg\ell_0$), as the standard expression is restored, so does the value of dimensions, which coincides with the conventional infrared dimension four. This method of identifying ultraviolet dimension and the result itself is similar to the conclusions drawn in \cite{arzano2017non}. Moreover, for extensive and important discussions on the relation between the spectral dimension and the propagator in quantum field theories, and in certain cases, the advantages of using propagators in relativistic QFTs to extract the the dimensional information over a non-relativistic diffusion equation with auxiliary time parameter, the readers are referred to \cite{modesto2012super,calcagni2016quantum}. We also note that it is also possible to calculate the explicit expression for the QG corrected potential in a general $D=d+1$ dimensional case, which we show in appendix \ref{Appendix E}.\par
Therefore, depending on whether we probe the short-distance structure of spacetime with massless or massive force carriers, the ultraviolet dimension of spacetime seems to be different while the infrared dimension coincides with the expected value in both cases.\par
\begin{figure}
    \centering
    \includegraphics[width=0.48\textwidth]{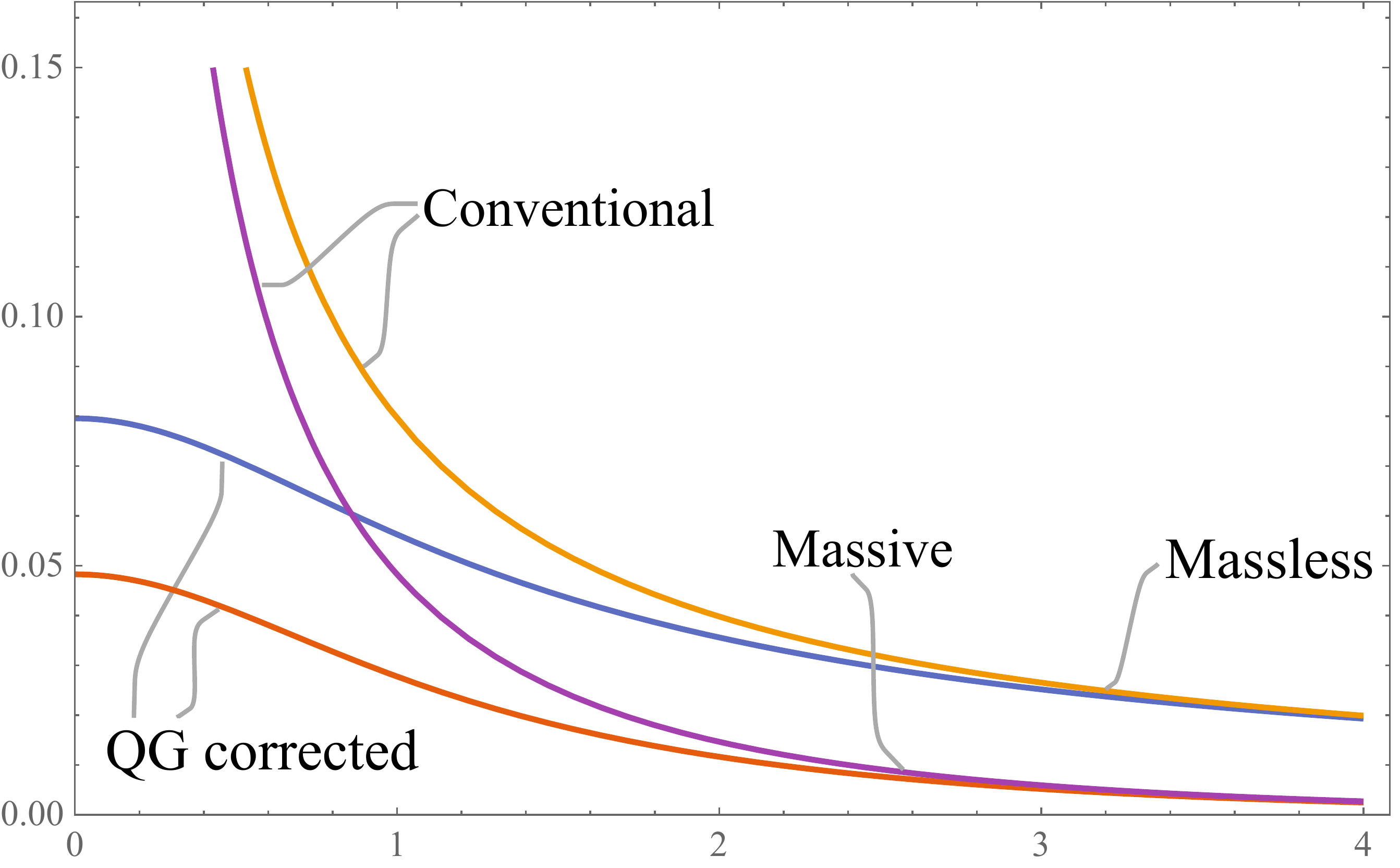}
    \caption{Plot of $-V(r)$ vs. $r$ for massive and massless carriers in the case of QG correction and their conventional counterparts. At $r=0$, the QG corrected potentials have finite value in contrast to the conventional potentials, which diverge.}
    \label{fig:my_label2}
\end{figure}

As the potential energy in the conventional case varies as $1/r$, in the limit $r\to 0$, it diverges. However, with the quantum gravity correction, as the zero-point length acts as a regulator, the potential energy tapers off to a finite value in the microscopic length scales (figure \ref{fig:my_label2}).

\section{Summary of results}
We have studied the dimensional reduction phenomena in various contexts. We have considered the spectral dimension, which near the scale of the zero point length reduces to $3.5$ from the value $4$, whereas, in the deep ultraviolet limit, the spectral dimension diverges. In the case of thermodynamic dimension, the Planckian value is $1.5$, and the deep ultraviolet value is $1$. The interaction potential dimension depends on the massive or massless nature of the force carriers producing the value $3$ and $2$, respectively, in the deep ultraviolet regime. Below we list these results for quick reference. 

\begin{table}[H]
\centering
\begin{tabular}{>{\arraybackslash}m{3.0cm}|>{\centering\arraybackslash}m{2cm}|>{\centering\arraybackslash}m{2cm}}
\hline
Method & $\sim$ Planck scale & Deep ultraviolet \\
\hline
Spectral & $3.5$ & $-\infty$\\\hline
    Thermodynamic & $1.5$ & $1$\\\hline
    Interaction potential (massive carrier) &  & $2$\\\hline
    Interaction potential (massless carrier) &  & $3$\\\hline
    Geometirc \cite{2016} & $2$ & \\\hline
\end{tabular}
\end{table}
We have borrowed the dimension computed via geometric arguments, that is, the dimension inferred from the scaling of the volume of a geodesic ball near the Planck scale, form \cite{2016}. Moreover, the equation of state parameter for a massless free field or ideal gas flows from the value $1/3$ in the infrared regime to a value $2$ near the Planck scale, signifying a continuous phase-transition.\par
Though the physical results derived here are similar to that of in \cite{2016}, an important distinction from our result should be noted. Whereas in \cite{2016}, the dimension of spacetime with zero-point length starting from a value $D$ in the infrared regime reduces at the Planck scale to $2$, we find that for spectral dimension and thermodynamic dimension, the values are $D-({1}/{2})$ and ${(D-1)}/{2}$, respectively. Thus, unlike the geometric dimension, the spectral and thermodynamic dimensions at the Planck scale depend on their values in the infrared limit. The interaction potential dimension in the ultraviolet limit, on the other hand, does not depend on its infrared value. The value for the thermodynamic dimension in the deep ultraviolet regime is independent of infrared value as well.
\section{Discussion}
It is reasonable to believe that the Planck length ($\ell_{\rm Pl}$) acts as a fundamental lower bound to the length scale of all physical processes. The hypothesis of path integral duality put forward by Padmanabhan proposes to include this bound by replacing the Euclidean geodesic distances $\sigma^2$ with $\sigma^2+\ell_0^2$. This operation leads to a factor of $e^{-{\ell_0^2}/{4s}}$ to be multiplied with the standard Schwinger heat kernel. The value of the zero-point length being extremely small, $\ell_0\propto\ell_{\rm Pl}$, its effects are not felt in the large distance scales. However, as one probes the mesoscopic regime near the Planck length, this factor becomes prevalent. When the auxiliary diffusion time goes below the zero-point length $s\ll \ell_0^2/4$, the exponential factor heavily suppresses the probability density of such diffusion processes taking place in smaller length scales than the Planck length. This suppression then can be a manifestation of reduced degrees of freedom in the ultraviolet regime due to dimensional reduction.\par
As there is no unique way to study dimensional reduction, we study this phenomenon in various contexts. We see that, although all the approaches indicate a reduction in dimensions in the short length scales, there is no unique number to which the dimension flows. This lack of uniqueness can be understood in the sense that we are probing the short-distance structure of spacetime by using different processes, i.e., diffusion, thermodynamic, and so on, whose behavior depends on the degrees of freedom allowed to these. Therefore, the difference in the precise value of the ultraviolet dimension indicates that the quantum \textit{fuzziness} of the spacetime does not screen dimensions for all processes equally. For example, in the case of massive force carriers, the dimensions get screened more heavily than for the case of massless force carriers resulting in a lower value for the ultraviolet dimension in the case of massive carriers. Moreover, one expects that the ultraviolet dimension of spacetime be 2, as in this case the quantum theory of gravity is expected to be renormalizable. As far as, our quantum field theoretic calculations are concerned, for massless force-carriers we see that the ultraviolet dimension does indeed reduce to the value 2, which also coincides with the value obtained from geometric arguments (scaling of the volume of a geodesic ball), suggesting this value to be the more reasonable one.\par
The mechanism by which the quantum gravity effects are being included is central to studying the ultraviolet nature of spacetime. For example, in \cite{modesto2010spectral} a minimum length of the spacetime has been incorporated by considering the smearing of delta distributions into a gaussian form of a width of the minimum length. In such a case, the results are different from our approach even though in both cases, the goal is to study the ultraviolet dimensional reduction in spacetime admitting a minimum length. Moreover, even though in most models of quantum gravity, the dimensional flow is typically from $4$ (infrared) to $2$ (ultraviolet), there are models in which the ultraviolet dimension does not even decrease but increases \cite{eichhorn2014spectral} or can diverge \cite{PhysRevD.102.086003,mandrysz2018ultralocal}.\par
We note that as the ideal gas in the zero-point length spacetime behaves like an exotic fluid having an effective equation of state parameter $w=2$ near the Planck scale, it can have interesting cosmological features. By recalling that the scale factor dependency of the energy density of cosmic fluids is given by $a^{-3(1+w)}$, near the Planck scale, the ideal gas's energy density scales as $a^{-9}$. As the obtained result is only valid near the Planck scale, it would be interesting to study the effects of this exotic fluid in the context of quantum cosmology in its effects on the initial singularity of the universe, as we see in the case of BKL-like instabilities \cite{belinskii1970oscillatory} caused by the presence of `stiff' matter or shear stress density. On the other hand, the potential of interaction between two point-like static classical sources asymptotically reaches a constant value; thus, the force between them grows progressively weaker in the short distances, which has a thought-provoking similarity to the asymptotic freedom in QCD \cite{gross1987asymptotic}.\par
Finally, the prescription to add a zero-point length to the Euclidean geodesic distances cannot be a standalone substitute for a quantum theory of gravity; instead, this result has to be embedded into a more sophisticated fundamental theory of the quantum nature of spacetime, which is expected to explain the mechanism of dimensional reduction adequately at short distance scales. For this reason, we do not trust or attempt to speculate on the nature of the anomalous results we have obtained---for example, the indeterminacy of the spectral dimension or the equation of state parameter in the deep ultraviolet limit---as the validity of the hypothesis might be limited to the mesoscopic scales or at best up to the Planck scale. Beyond that scale, the fundamental theory must take over.
\begin{acknowledgements}
We gratefully acknowledge several helpful comments from Sumanta Chakraborty and Karthik Rajeev, which have improved the content of this work. This research is funded by the INSPIRE fellowship from the DST, Government of India (Reg. No. DST/INSPIRE/03/2019/001887).
\end{acknowledgements}

\appendix
\section{Gamma-type integration}\label{appendix A}
We start with
\begin{align}
    \int_0^\infty {\rm d}y \, y^{\frac{D}{2}-1}e^{-y(n^2+\alpha^2)}.
\end{align}
Substituting $z=y(n^2+\alpha^2)$, reduces this integral into the standard gamma function integration. However, a more general technique exists for Mellin transformations.\par Expanding the exponential we can write the above integral as
\begin{align}
    \int_0^\infty {\rm d}y \, y^{\frac{D}{2}-1}\sum_{m=0}^{\infty}\frac{(-y)^m (n^2+\alpha^2)^m}{m!}.
\end{align}
Ramanujan's master theorem states that for functions that can be expressed as
\begin{align}
    f(x)=\sum_{m=0}^\infty\frac{\varphi(m)}{m!}(-x)^m,
\end{align}
their Mellin transformation results into
\begin{align}
    \int_0^\infty x^{s-1}f(x){\rm d}x=\Gamma(s)\varphi(-s).
\end{align}
Identifying $\varphi(m)$ with $(n^2+\alpha^2)^m$ and $s$ with $\frac{D}{2}$, we get
\begin{align}
    \int_0^\infty {\rm d}y \, y^{\frac{D}{2}-1}e^{-y(n^2+\alpha^2)}=\Gamma\left(\frac{D}{2}\right)\frac{1}{\left(n^2+\alpha^2\right)^{\frac{D}{2}}}.
\end{align}
Thus we have obtained the desired result.
\section{Integral representation of the generalized Mathieu series}\label{Appendix B}
We are dealing wih a generalized Mathieu series of the following form
\begin{equation}
    \sum_{n=1}^{\infty}\frac{1}{(n^2+\alpha^2)^{\frac{D}{2}}}.
\end{equation}
The above series can be expressed in terms of hypergeometric function $F(a,b;c;z)$ using the following identity (Eq. (9.121.1) in \cite{gradshte2007table})
\begin{equation}
    F\left(-a,b;b;-z\right)=(1+z)^{a},
\end{equation}
Thus we have
\begin{align}
    &\sum_{n=1}^{\infty}\frac{n^D}{(n^2+\alpha^2)^{\frac{D}{2}}}=\sum_{n=1}^\infty F\left(\frac{D}{2},\frac{D+1}{2};\frac{D+1}{2};-\frac{\alpha^2}{n^2}\right).
\end{align}
Then we can use the identity (6.621.1) in \cite{gradshte2007table}
\begin{align}
    \int_0^\infty t^{\mu-1}e^{-at}J_\nu(bt){\rm d}t=\frac{\left(\frac{1}{2}b\right)^\nu}{a^{\mu+\nu}}\frac{\Gamma(\mu+\nu)}{\Gamma(\nu+1)}\nonumber\\
    \times F\left(\frac{\mu+\nu}{2},\frac{\mu+\nu+1}{2};\nu+1;-\frac{b^2}{a^2}\right).
\end{align}
Identifying
\begin{align}
    \mu=\frac{D+1}{2},\quad \nu=\frac{D-1}{2},\quad b=\alpha, \quad a=n,
\end{align}
we immediately get
\begin{align}
    \sum_{n=1}^{\infty}\frac{1}{(n^2+\alpha^2)^{\frac{D}{2}}}=\frac{\sqrt{\pi}}{\Gamma\left(\frac{D}{2}\right)}\left(2\alpha\right)^{-\frac{D-1}{2}}\nonumber\\
    \times\sum_{n=1}^\infty\int_0^\infty x^{\frac{D-1}{2}}e^{-nx}J_{\frac{D-1}{2}}(\alpha x){\rm d}x,
\end{align}
where we have used the fact that
\begin{align}
    \Gamma\left(\frac{D}{2}\right)\Gamma\left(\frac{D+1}{2}\right)=2^{1-D}\sqrt{\pi}\Gamma\left(D\right).
\end{align}
Using Fubini-Tonelli's theorem, we can pass the summation through the integration to get
\begin{align}
    \sum_{n=1}^{\infty}\frac{1}{(n^2+\alpha^2)^{\frac{D}{2}}}=\frac{\sqrt{\pi}}{\Gamma\left(\frac{D}{2}\right)}\left(2\alpha\right)^{-\frac{D-1}{2}}\nonumber\\
    \times\int_0^\infty \frac{x^{\frac{D-1}{2}} J_{\frac{D-1}{2}}(\alpha x)}{e^x-1}{\rm d}x.
\end{align}
This is the expression we were after. For further details on the generalized Mathieu series see \cite{tomovski2007integral}.
\section{Asymptotic expansion of the Hankel transform}\label{Appendix C}
The identity for Bernoulli numbers (Sec. (9.610) \cite{gradshte2007table})
\begin{align}
    \frac{x}{e^x-1}=\sum_{k=0}^\infty B_k\frac{x^k}{k!}
\end{align}
implies
\begin{align}
    \frac{x^\nu}{e^x-1}=\sum_{k=0}^\infty B_k\frac{x^{k+\nu-1}}{k!}.
\end{align}
Now the using the following integration formula for Bessel function (see Eq.~(6.561.14) in \cite{gradshte2007table})
\begin{align}
    \int_0^\infty x^\mu J_\nu(\alpha x){\rm d}x=\frac{1}{2}\left(\frac{2}{\alpha}\right)^{\mu+1}\frac{\Gamma\left(\frac{1}{2}\mu+\frac{1}{2}\nu+\frac{1}{2}\right)}{\Gamma\left(\frac{1}{2}\nu-\frac{1}{2}\mu+\frac{1}{2}\right)},
\end{align}
(with $-{\rm Re}\,\nu-1<{\rm Re}\,\mu<\frac{1}{2},$ and $\alpha>0$) we can approximately evaluate the Hankel transform for large argument limit (see theorem 2 in \cite{wong1976error}) as 
\begin{align}
    \int_0^\infty\frac{x^\nu J_{\nu}(\alpha x)}{e^x-1}{\rm d}x&\simeq\frac{1}{2}\sum_{k=0}^\infty \frac{B_k}{k!}\left(\frac{2}{\alpha}\right)^{k+\nu}\frac{\Gamma\left(\nu+\frac{1}{2}k\right)}{\Gamma\left(1-\frac{1}{2}k\right)}\nonumber\\
    &= \frac{\Gamma(\nu)}{2}\left(\frac{2}{\alpha}\right)^{\nu}-\frac{\Gamma(\nu+\frac{1}{2})}{4\sqrt{\pi}}\left(\frac{2}{\alpha}\right)^{1+\nu}
\end{align}
where in the last line we have recalled the fact that for odd numbers $k\geq 3$, the Bernoulli's numbers are zero and for even numbers $k> 3$ the reciprocal gamma function vanishes as $\frac{1}{\Gamma(-m)}=0$ for $m=0,1,2,3,\dots$. Therefore, all the other terms in the expansion vanishes except for the first two terms.
\section{Yukawa potential in \texorpdfstring{$d$}{d} dimensions}\label{Appendix D}
The Yukawa potential in $d$ dimension has the following integral form
\begin{align}
    V_{d+1,\,{\rm Yukawa}}(r)&=-\int \frac{{\rm d}^d \mathbf{p}}{(2\pi)^d}\frac{e^{i\mathbf{p}\cdot\mathbf{r}}}{|\mathbf{p}|^2+m^2}\nonumber\\
    &=-\frac{S_{d-1}}{(2\pi)^d}\int_0^\infty |\mathbf{p}|^{d-1}({\rm d}|\mathbf{p}|)\nonumber\\
    &\hspace{2cm}\int_0^\pi{\rm d}\theta \sin^{d-2}\theta\frac{e^{i|\mathbf{p}|r\cos \theta}}{|\mathbf{p}|^2+m^2},
\end{align}
where $S_{d-1}$ is the respective solid angle. Let us focus on the integration first
\begin{align}
    &\int_0^\infty |\mathbf{p}|^{d-1}({\rm d}|\mathbf{p}|)\int_0^\pi {\rm d}\theta\sin^{d-2}\theta\frac{e^{i|\mathbf{p}|r\cos \theta}}{|\mathbf{p}|^2+m^2}\nonumber\\
    &=\int_0^\infty |\mathbf{p}|^{d-1}({\rm d}|\mathbf{p}|)\frac{\sqrt{\pi}\Gamma\left(\frac{d-1}{2}\right)\left(\frac{i|\mathbf{p}|r}{2}\right)^{\frac{2-d}{2}}I_{\frac{d-2}{2}}(i|\mathbf{p}|r)}{|\mathbf{p}|^2+m^2}\nonumber\\
    &=\frac{\sqrt{\pi}\Gamma\left(\frac{d-1}{2}\right)}{\left(\frac{1}{2}r\right)^{\frac{d-2}{2}}}\int_0^\infty({\rm d}|\mathbf{p}|) |\mathbf{p}|^{\frac{d}{2}}\frac{J_{\frac{d-2}{2}}(|\mathbf{p}|r)}{|\mathbf{p}|^2+m^2}
\end{align}
Here, for the first equality, we have used the integral representation of $I_{\nu}(z)$ (see, Eq.~(8.431.3) \cite{gradshte2007table})
\begin{align}
    I_{\nu}(z)=\frac{\left(\frac{z}{2}\right)^\nu}{\Gamma\left(\nu+\frac{1}{2}\right)\Gamma\left(\frac{1}{2}\right)}\int_0^\pi e^{\pm z \cos\theta}\sin^{2\nu}\theta {\rm d}\theta,
\end{align}
and for the second equality we have used the relations
\begin{align}
    I_{\nu}(z)=i^{-\nu}J_{\nu}(iz);\qquad J_{\nu}(-z)=e^{i\pi\nu}J_{\nu}(z).
\end{align}
The rest of the integration can be done using the formula Eq.~(6.565.4) in \cite{gradshte2007table}
\begin{align}\label{eq:good_integral}
    \int_0^\infty({\rm d}|\mathbf{p}|) |\mathbf{p}|^{\frac{d}{2}}\frac{J_{\frac{d-2}{2}}(|\mathbf{p}|r)}{|\mathbf{p}|^2+m^2}={m^{\frac{d-2}{2}}}K_{\frac{d-2}{2}}(mr).
\end{align}
Taking care of all the prefactors, and using the relation for the solid angle
\begin{align}
    S_{d-1}=\frac{2\pi^{\frac{d-1}{2}}}{\Gamma\left(\frac{d-1}{2}\right)}
\end{align}
we get the following expression for the potential energy
\begin{align}
    V_{d+1,\,{\rm Yukawa}}(r)=-\frac{1}{(2\pi)^{\frac{d}{2}}}\left(\frac{m}{r}\right)^{\frac{d-2}{2}}K_{\frac{d-2}{2}}(mr)
\end{align}
For massless case the integration (\ref{eq:good_integral}) can be performed using the formula Eq.~(6.561.14) in \cite{gradshte2007table}
\begin{align}
    \int_0^\infty({\rm d}|\mathbf{p}|) |\mathbf{p}|^{\frac{d}{2}-2}{J_{\frac{d-2}{2}}(|\mathbf{p}|r)}=\frac{2^{\frac{d}{2}-2}}{r^{\frac{d-2}{2}}}{\Gamma\left(\frac{d-2}{2}\right)}.
\end{align}
Therefore, the potential energy in the massless case is
\begin{align}
    V_{d+1,\,{\rm Yukawa,}\, m=0}(r)=-\frac{1}{4\pi^{\frac{d}{2}}}\left(\frac{1}{r}\right)^{d-2}\Gamma\left(\frac{d-2}{2}\right).
\end{align}
\section{QG correction in \texorpdfstring{$D$}{D} dimensions}\label{Appendix E}
The QG corrected potential energy in the $D=d+1$ dimensions has the following integral form
\begin{align}
    V^{\ell_0}_{d+1}(r)&=-\int \frac{{\rm d}^d \mathbf{p}}{(2\pi)^d}\frac{e^{i\mathbf{p}\cdot\mathbf{r}}\ell_0K_{1}\left(\ell_0\sqrt{|\mathbf{p}|^2+m^2}\right)}{\sqrt{|\mathbf{p}|^2+m^2}}\nonumber\\
    &=-\frac{S_{d-1}}{(2\pi)^d}\int_0^\infty |\mathbf{p}|^{d-1}({\rm d}|\mathbf{p}|)\nonumber\\
    &\int_0^\pi{\rm d}\theta \sin^{d-2}\theta\frac{e^{i|\mathbf{p}|r\cos \theta}\ell_0 K_{1}\left(\ell_0\sqrt{|\mathbf{p}|^2+m^2}\right)}{\sqrt{|\mathbf{p}|^2+m^2}}.
\end{align}
Focusing on the integration part first, we get
\begin{align}
    &\int_0^\infty |\mathbf{p}|^{d-1}({\rm d}|\mathbf{p}|)\int_0^\pi{\rm d}\theta \sin^{d-2}\theta e^{i|\mathbf{p}|r\cos \theta}\nonumber\\
    &\hspace{2cm}\times\frac{\ell_0 K_{1}\left(\ell_0\sqrt{|\mathbf{p}|^2+m^2}\right)}{\sqrt{|\mathbf{p}|^2+m^2}}\nonumber\\
    &=\int_0^\infty |\mathbf{p}|^{d-1}({\rm d}|\mathbf{p}|){\sqrt{\pi}\Gamma\left(\frac{d-1}{2}\right)\left(\frac{i|\mathbf{p}|r}{2}\right)^{\frac{2-d}{2}}}\nonumber\\
    &\hspace{2cm}\times I_{\frac{d-2}{2}}(i|\mathbf{p}|r)\frac{\ell_0 K_{1}\left(\ell_0\sqrt{|\mathbf{p}|^2+m^2}\right)}{\sqrt{|\mathbf{p}|^2+m^2}}\nonumber\\
    &=\frac{\sqrt{\pi}\Gamma\left(\frac{d-1}{2}\right)}{\left(\frac{1}{2}r\right)^{\frac{d-2}{2}}}\int_0^\infty({\rm d}|\mathbf{p}|) |\mathbf{p}|^{\frac{d}{2}}{J_{\frac{d-2}{2}}(|\mathbf{p}|r)}\nonumber\\
    &\hspace{2cm}\times\frac{\ell_0 K_{1}\left(\ell_0\sqrt{|\mathbf{p}|^2+m^2}\right)}{\sqrt{|\mathbf{p}|^2+m^2}},
\end{align}

where again we have used the integral representation of $I_{\nu}(x)$ and its imaginary argument relations. The rest of the integration can be done using the formula Eq.~(6.596.7) in \cite{gradshte2007table}
\begin{align}\label{eq:good_integral 2}
    \int_0^\infty({\rm d}|\mathbf{p}|) |\mathbf{p}|^{\frac{d}{2}}\frac{J_{\frac{d-2}{2}}(|\mathbf{p}|r){\ell_0 K_{1}\left(\ell_0\sqrt{|\mathbf{p}|^2+m^2}\right)}}{{\sqrt{|\mathbf{p}|^2+m^2}}}\nonumber\\
    =\left(\frac{mr}{\sqrt{r^2+\ell_0^2}}\right)^{\frac{d-2}{2}}K_{-\frac{d-2}{2}}\left(m\sqrt{r^2+\ell_0^2}\right).
\end{align}
Taking care of all the prefactors we get
\begin{align}
    V^{\ell_0}_{d+1}(r)=-\frac{1}{(2\pi)^{\frac{d}{2}}}\left(\frac{m}{\sqrt{r^2+\ell_0^2}}\right)^{\frac{d-2}{2}}K_{-\frac{d-2}{2}}\left(m\sqrt{r^2+\ell_0^2}\right).
\end{align}
For massless case the integration (\ref{eq:good_integral 2}) can be performed using the formula Eq.~(6.576.3) \cite{gradshte2007table}
\begin{align}
    &\int_0^\infty({\rm d}|\mathbf{p}|) |\mathbf{p}|^{\frac{d}{2}-1}{J_{\frac{d-2}{2}}(|\mathbf{p}|r)}\ell_0 K_{1}\left(\ell_0|\mathbf{p}|\right)\nonumber\\
    &=\frac{r^{\frac{d-2}{2}}\Gamma\left(\frac{d-2}{2}\right)}{2^{2-\frac{d}{2}}\ell_0^{d-2}}F\left(\frac{d}{2},\frac{d-2}{2};\frac{d}{2};-\frac{r^2}{\ell_0^2}\right)\nonumber\\
    &=\frac{r^{\frac{d-2}{2}}\Gamma\left(\frac{d-2}{2}\right)}{2^{2-\frac{d}{2}}}\left({r^2+\ell_0^2}\right)^{-\frac{d-2}{2}}.
\end{align}

Therefore, the potential energy in the massless case reads
\begin{align}
    V^{\ell_0}_{d+1,\,m=0}(r)=-\frac{\Gamma\left(\frac{d-2}{2}\right)}{4\pi^{\frac{d}{2}}}\left(r^2+\ell_0^2\right)^{-\frac{d-2}{2}}.
\end{align}
\bibliographystyle{apsrev4-2}
\bibliography{reference}
\end{document}